\documentclass[english]{article}
\PassOptionsToPackage{natbib=true}{biblatex}
\usepackage[latin9]{inputenc}
\usepackage{geometry}
\usepackage[active]{srcltx}
\usepackage{float}
\usepackage{textcomp}
\usepackage{amsmath}
\usepackage{amssymb}
\usepackage{stmaryrd}
\usepackage{graphicx}
\usepackage{setspace}
\makeatletter
\newcommand{\lyxaddress}[1]{
	\par {\raggedright #1
	\vspace{1.4em}
	\noindent\par}
}

\@ifundefined{date}{}{\date{}}
\makeatother

\usepackage{babel}
\usepackage[bibstyle=numeric,backend=bibtex]{biblatex}
\begin{document}
\title{A possibility of Klein Paradox in quaternionic ($3+1$) frame}
\author{Geetanjali Pathak and B. C. Chanyal\textbf{}\thanks{Corresponding author (email: bcchanyal@gmail.com)}}
\maketitle

\lyxaddress{\begin{center}
\textit{Department of Physics}\\
\textit{G. B. Pant University of Agriculture and Technology, Pantnagar
263 145, Uttarakhand, India}\\
\par\end{center}}
\begin{abstract}
In light of the significance of non-commutative quaternionic algebra
in modern physics, the current study proposes the existence of the
Klein paradox in the quaternionic (3+1)-dimensional space-time structure.
By introducing quaternionic wave-function, we rewrite the Klein-Gordon
equation in extended quaternionic form that includes scalar and the
vector fields. Because quaternionic fields are non-commutative, the
quaternionic Klein-Gordon equation provides three separate sets of
the probability density and probability current density of relativistic
particles. We explore the significance of these probability densities
by determining the reflection and transmission coefficients for the
quaternionic relativistic step potential. Furthermore, we also discuss
the quaternionic version of the oscillatory, tunnelling, and Klein
zones for the quaternionic step potential. The Klein paradox occurs
only in the Klein zone when the impacted particle's kinetic energy
is less than $\mathbb{V}_{0}-m_{0}c^{2}$. Therefore, it is emphasized
that for the quaternionic Klein paradox, the quaternionic reflection
coefficient becomes exclusively higher than value one while the quaternionic
transmission coefficient becomes lower than zero.

\end{abstract}

\section{Introduction}

For the development of the microscopic world in terms of hypercomplex
division algebra, quaternionic quantum mechanics (qQM) has been mathematically
developed as a modified version of quantum theory that uses typical
complex algebra. The quaternions {[}1{]} are similar to complex numbers
in values, but their multiplication is non-commutative, giving rise
to some additional degree of freedom. The qQM is formulated using
hypercomplex wave functions {[}2{]} evaluated over the quaternionic
number field. The quaternions were utilized in discussions of a novel
area of quantum physics {[}3{]}, where the quaternionic field characterizes
the reducibility of finite dimension. Adler {[}4{]} has researched
and compiled information on the quaternionic generalization of quantum
mechanics and quantum field theory. Since the quaternion structure
offers homogeneous space-time in higher-dimensional quantum mechanics
and has the potential to unify all four fundamental interactions,
it has replaced the space-time formalism {[}5{]} as the dominant space-time
structure in theoretical physics. Therefore, quaternions are well
capable of dealing with higher dimensions. Furthermore, the quaternions
are substantially more compact than the standard form of Lorentz vectors
and tensors for any solution that is conceivable. Also, quaternionic
units are used to define the Pauli spin {[}2,6{]} corresponding to
the spin of the particle. As a part of this, the quaternions were
effectively applied in the formulation of a theory for the Dirac equations,
which disclosed various aspects of employing quaternions in relativistic
quantum mechanics {[}7{]}. Giardino {[}8-10{]} studied the various
applications of quaternionic quantum mechanics by using the concept
of real Hilbert space. The formulation and resolution of the quaternionic
form of the Klein-Gordon equation were the focus of Giardino\textquoteright s
work {[}11{]} whereas Seema {[}12{]} provided a pure quaternionic
Klein-Gordon equation and quaternionic wave. The additional parameters
of this non-commutative algebra facilitate the investigation of concealed
physical states in complicated circumstances. As such, the Dirac equation
has been derived in quaternionic space {[}13,14{]} and the subject
also concerns the formulation of the quaternionic Klein-Gordon equation
for a scalar field {[}11,12,15{]}. Not only this, many researchers
{[}16-28{]} have used the quaternion algebra for applications in the
various disciplines of physics.

In relativistic qQM, it is very challenging to consider four-dimensional
relativistic quaternionic potential with the application of Klein-Gordon
and Dirac field equations. Initially, Leo et al. {[}29{]} provided
the key ideas of quaternionic value potential for solving the quaternionic
Dirac equation. Further, the quantum tunnelling phenomena {[}30,31{]}
for relativistic fermions have been studied in the presence of a quaternionic
step potential. In this context, the Klein-paradox {[}32{]} is an
interesting challenge to discuss in relativistic qQM, which shows
the anomalous reflection of particles off a huge potential barrier.
Many physicists {[}33-35{]} worked on the Klein-paradox for relativistic
quantum mechanics. Recently, the role of quaternionic four-spaces
in the fundamental of relativistic quantum mechanics has been explored
{[}36,37{]}. However, the Klein-paradox problem can be solved by using
relativistic qQM, which can provide some interesting aspects. Therefore,
in this way, considering the non-commutative nature of quaternions
under multiplication operation and its significance in relativistic
qQM, we have used quaternionic four spaces, i.e. ($3+1$) spaces (where
three spaces for spatial coordinate and one space component for temporal
coordinate), to discuss Klein-Gordon equation for scalar as well as
the vector field by using quaternionic wave function. In the quaternionic
field, we propose three different cases to discuss the probability
density and probability current density of relativistic particles,
where one case is precisely matched for a pure scalar field, the second
case corresponds to a pure vector field, and the third case is applicable
to a combined field containing both the scalar and vector fields.
We further define the relativistic step-potential in quaternionic
($3+1$) spaces to analyze the application of the quaternionic Klein-Gordon
equation. We determine the reflection and transmission coefficients
for a quaternionic plane wavefunction interacting with a quaternionic
relativistic step potential. We offer three separate energy zones
related to particle's momentum to address different values of the kinetic
energy of particles. As a result, the quaternionic form of the Klein
paradox is established for the proposed Klein zone\textbf{ }where
the kinetic energy $E$ of incident particle is less than $\mathbb{V}_{0}-m_{0}c^{2}$.
In the foregoing results, it has been proposed that the Klein paradox
occurs when the momentum direction of particle in step-potential region
is opposite to its incident momentum direction.

\section{Preliminaries}

A quaternionic field is a kind of generalized field made up of four-dimensional
spaces that include respectively, scalar and vector fields. Thus,
a quaternion variable $\mathbb{Q}$ can be expressed as
\begin{align}
\mathbb{Q}\,:\longmapsto\,\,\,\mathbb{Q}(e_{0},e_{j})\,= & \,\,e_{0}q_{0}+\sum_{j=1}^{3}e_{j}q_{j}\,\,,\,\,\,\,\forall\,\,q_{0}\in\mathbb{R},\,q_{j}\in\mathbb{R}^{3}\,,\label{eq:1}
\end{align}
where ($e_{0},e_{1},e_{2},e_{3}$) are the quaternionic basis elements
in which $e_{0}$ denotes a scalar unit and $e_{j}$ $\left(\text{for}\,j=1,2,3\right)$
denotes the vector unit. The quaternionic conjugate of variable $\mathbb{Q}$,
defined by $\mathbb{Q}^{*}$, can be written as
\begin{align}
\mathbb{Q^{*}}\,:\longmapsto\,\,\,\mathbb{Q}^{*}(e_{0},-e_{j})\,= & \,\,e_{0}q_{0}-\sum_{j=1}^{3}e_{j}q_{j}\,\,.\label{eq:2}
\end{align}
The quaternion conjugation is an involution or its own inverse, it
always gives the original element when applied to an element twice,
i.e., $(\mathbb{Q}^{*})^{*}=\mathbb{Q}$. The quaternion conjugate
of any two quaternionic variable can also be written as $(\mathbb{Q}_{1}\mathbb{Q}_{2})^{*}=\mathbb{Q}_{2}^{*}\mathbb{Q}_{1}^{*}$.
However, because the quaternionic multiplication $(\mathbb{Q}_{1}\mathbb{Q}_{2})$
depends upon the multiplication of their basis elements, one can write
the multiplication relations of quaternion units as
\begin{align}
e_{0}^{2}\,= & \,1,\,\,\,e_{j}^{2}\,=-1,\,\,\,\,e_{0}^{*}\,=\,e_{0},\,\,\,\,e_{j}^{*}\,=\,-e_{j}\,,\nonumber \\
e_{0}e_{j}\,= & \,e_{j}e_{0}\,=\,e_{j},\,\,\,e_{i}e_{j}\,=-\delta_{ij}+\epsilon_{ijk}e_{k},\,\,\,\forall\,(i,j,k=1,2,3),\label{eq:3}
\end{align}
where $\delta_{ij}$ and $\epsilon_{ijk}$ are respectively, the Kronecker
delta and the three index Levi-Civita symbols. Therefore, from Eq.(3),
the multiplication of two quaternions $(\mathbb{Q}_{1}\mathbb{Q}_{2})$
can be employed as a non commutative quaternionic field as follows:
\begin{align}
\mathbb{Q}_{1}\mathbb{Q}_{2}\,= & \,\,e_{0}(q_{0}q'_{0}-\boldsymbol{q}.\boldsymbol{q}')+\sum_{j=1}^{3}\,e_{j}\left[q_{0}\boldsymbol{q}'+q'_{0}\boldsymbol{q}+(\boldsymbol{q}\times\boldsymbol{q}')_{j}\right]\,.\label{eq:4}
\end{align}
In this situation, the coefficient $e_{0}$ is merely a scalar quantity
that may be stated in the presence of a quaternionic scalar field,
but the coefficient $e_{j}$ can be applied to a pure quaternionic
vector field. Thus, the quaternions show non-commutativity under multiplication
operation as $\mathbb{Q}_{1}\mathbb{Q}_{2}\neq\mathbb{Q}_{2}\mathbb{Q}_{1}$
because of $(\boldsymbol{q}\times\boldsymbol{q}')_{j}\neq(\boldsymbol{q}'\times\boldsymbol{q})_{j}$.
The quaternionic multiplication of two quaternions may exhibit a commutative
nature only if the vector products of their vectors $\boldsymbol{q}$
and $\boldsymbol{q}'$ are zero or $\boldsymbol{q}$ is parallel to
$\boldsymbol{q}'$. The factor $\left(\boldsymbol{q}\times\boldsymbol{q}'\right)$
plays a significant role for vector field quantities like spin angular
momentum, orbital angular momentum, etc. Additionally, quaternionic
fields are capable of having the associative property under multiplication,
such as $\mathbb{Q}_{1}\left(\mathbb{Q}_{2}\mathbb{Q}_{3}\right)=\left(\mathbb{Q}_{1}\mathbb{Q}_{2}\right)\mathbb{Q}_{3}.$
The norm of a quaternion is the square root of the product of the
quaternion and its conjugate as
\begin{align}
\mathbb{\left\Vert Q\right\Vert \,=} & \,\,\sqrt{q_{0}^{2}+q_{1}^{2}+q_{2}^{2}+q_{4}^{3}}\,\,=\,\,\sqrt{\mathbb{Q}\mathbb{Q}^{*}}\,\,=\,\,\sqrt{\mathbb{Q^{*}}\mathbb{Q}}\,,\label{eq:5}
\end{align}
which is always a real number and cannot be negative. The quaternionic
norm of any two variables shows the multiplicative property, which
means $\left\Vert \mathbb{Q}_{1}\mathbb{Q}_{2}\right\Vert =\left\Vert \mathbb{Q}_{2}\right\Vert \left\Vert \mathbb{Q}_{1}\right\Vert $.
In quaternionic division algebra, any nonzero quaternion has an inverse
with regard to the Hamilton product, that is,
\begin{align}
\left[e_{0}q_{0}+\sum_{j=1}^{3}e_{j}q_{j}\right]^{-1}\,\,= & \,\,\frac{1}{\sqrt{q_{0}^{2}+q_{1}^{2}+q_{2}^{2}+q_{4}^{3}}}\left[e_{0}q_{0}-\sum_{j=1}^{3}e_{j}q_{j}\right]\,.\label{eq:6}
\end{align}
Moreover, in the quaternionic metric space, the norm makes it possible
to define the distance $\mathbb{D}(\mathbb{Q}_{1},\,\mathbb{Q}_{2})$
between $\mathbb{Q}_{1}$ and $\mathbb{Q}_{2}$ by the norm of their
difference as $\mathbb{D}(\mathbb{Q}_{1},\,\mathbb{Q}_{2})=\left\Vert \mathbb{Q}_{1}-\mathbb{Q}_{2}\right\Vert $.
In terms of the corresponding metric topology, addition and multiplication
are continuous operations.

\section{Generalized quaternionic wave-function}

In modern quantum physics, a quaternionic wave function is a generalized
wave function that characterizes the quantum state mathematically.
Thus, the quaternionic wave-function is somewhat comparable to four
space-time wave-functions that involve four-dimensional non-commutative
quaternion algebra. Now, in order to write the quaternionic wave function,
let us start with quaternionic four$-$space ($\mathbb{X}$) and four$-$linear
momentum ($\text{\ensuremath{\mathbb{P}}}$) as
\begin{align}
\mathbb{X}(e_{0},e_{j})\,\,= & \,\,e_{0}x_{0}+\sum_{j=1}^{3}e_{j}x_{j}\,\,,\,\,\,\,\forall\,\,x_{0}\in\mathbb{R},\,x_{j}\in\mathbb{R}^{3}\,,\label{eq:7}\\
\mathbb{P}(e_{0},e_{j})\,\,= & \,\,e_{0}p_{0}+\sum_{j=1}^{3}e_{j}p_{j}\,\,,\,\,\,\,\forall\,\,p_{0}\in\mathbb{R},\,p_{j}\in\mathbb{R}^{3}\,,\label{eq:8}
\end{align}
where the coefficient of quaternionic scalar unit ($e_{0}$) can be
written as time component while the vector unit ($e_{j}$) can be
written as three-space components. In quaternionic Euclidean structure
($-,+,+,+$), Eqs. (7) and (8) can be rewrite as $\mathbb{X}(e_{0},e_{j})=\left(-ict,\,\boldsymbol{x}\right)$
and $\mathbb{P}(e_{0},e_{j})=\left(-iE/c,\,\boldsymbol{p}\right)$;
where $c$ is the speed of light. It should be noted that the quaternionic
structure has $(1_{\text{time}}+3_{\text{space}})$ or ($3_{\text{space}}+1_{\text{time}}$)
dimensions, where one dimension is linked to a scalar (time-like)
component and the other three are linked to vector (space-like) components.
Therefore, the quaternionic wave function or a wave state ($\Psi$)
can be generalized as, {[}12{]}
\begin{align}
\Psi(e_{0},e_{j})\,= & \,\,e_{0}\Psi_{0}\exp\left\{ i\Theta_{0}\right\} +\sum_{j=1}^{3}e_{j}\Psi_{j}\exp\left\{ i\Theta_{j}\right\} \,\,,\label{eq:9}
\end{align}
where $(\Psi_{0},\Psi_{j})$ are the scalar and vector parts of quaternionic
state vector while $\left\{ \Theta_{0},\Theta_{j}\right\} $ are the
corresponding quaternionic phase terms. In the superposition quantum
state, we have both scalar and vector states. For pure quaternionic
scalar state we can write $\Psi(e_{0},0)\,=\,e_{0}\Psi_{0}\exp\left\{ i\Theta_{0}\right\} $,
and for pure vector state, $\Psi(0,e_{j})\,=\,\sum_{j=1}^{3}e_{j}\Psi_{j}\exp\left\{ i\Theta_{j}\right\} $.
The quaternionic phase ($\Theta$) can be employed by the product
of conjugate four-momentum and four-space as
\begin{align}
\Theta(e_{0},e_{j})\,\,:\longmapsto\frac{1}{\hbar}\left(\mathbb{P}^{*}\mathbb{X}\right) & \,=\,\,\frac{1}{\hbar}\left[\left(e_{0}p_{0}-\sum_{j=1}^{3}e_{j}p_{j}\right)\left(e_{0}x_{0}+\sum_{j=1}^{3}e_{j}x_{j}\right)\right]\nonumber \\
 & =\,\,e_{0}\Theta_{0}+\sum_{j=1}^{3}e_{j}\Theta_{j}\,,\label{eq:10}
\end{align}
along with quaternionic values
\begin{align}
\Theta_{0}\,= & \,\,\frac{1}{\hbar}\left(p_{0}x_{0}+\boldsymbol{p}\cdot\boldsymbol{x}\right)\,=\,\,\frac{1}{\hbar}\left(-Et+\boldsymbol{p}\cdot\boldsymbol{x}\right)\,,\label{eq:11}\\
\Theta_{j}\,= & \,\,\frac{1}{\hbar}\left[x_{0}p_{j}+p_{0}x_{j}+\left(\boldsymbol{p}\times\boldsymbol{x}\right)_{j}\right]\,=\,\,\frac{i}{\hbar}\left[ct\boldsymbol{p}_{j}-\left(\frac{E}{c}\right)\boldsymbol{x}_{j}-i\left(\boldsymbol{p}\times\boldsymbol{x}\right)_{j}\right]\,.\label{eq:12}
\end{align}
Therefore, from Eq.(9) we gets
\begin{align}
\Psi(e_{0},e_{j})\,= & \,\,e_{0}\Psi_{0}\exp\left\{ -\frac{i}{\hbar}\left(Et-\boldsymbol{p}\cdot\boldsymbol{x}\right)\right\} +\sum_{j=1}^{3}e_{j}\Psi_{j}\exp\left\{ -\frac{1}{\hbar}\left[ct\boldsymbol{p}_{j}-\left(\frac{E}{c}\right)\boldsymbol{x}_{j}-i\left(\boldsymbol{p}\times\boldsymbol{x}\right)_{j}\right]\right\} \,\,.\label{eq:13}
\end{align}
Eq.(13) represents a generalized wave state of a particle in terms
of quaternionic basis. Although the real part of the wave state $\Psi(e_{0},0)$
has the conventional significance of representing the dynamics of
particle behavior, while the imaginary part of the quaternionic wave-state
$\Psi(0,e_{j})$ illustrates the damping nature along the $e_{j}-$axes.
In the four-dimensional quaternionic space, the attenuation of the
wave state will always be influenced by the pure quaternionic component.
Additionally, it is straightforward to write the probability of a
particle including both quaternionic components as $\mathcal{P}\sim\left|\mathcal{P}_{1}-\mathcal{P}_{2}\right|$
$=\left|\Psi_{0}\right|^{2}-\left|\Psi_{j}\right|^{2}\exp\left\{ -2z\right\} $
where $z=\frac{1}{\hbar c}\left(c^{2}tp_{j}-Ex_{j}\right)$.

\section{Quaternionic version of Klein-Gordon equation}

In order to describe the wave behavior of the particle in a quaternionic
field, we state the Klein-Gordon equation. Thus, let us begin with
the quaternionic eigenvalue equation for the quaternionic momentum
of the particle as
\begin{align}
\mathbb{P}\,\Psi\,= & \,\,m_{0}c\,\Psi\,.\label{eq:14}
\end{align}
For the second-order differential equation, we operate complex conjugate
of quaternionic momentum $\mathbb{P}^{*}$ by left to the both side
in the Eq.(14) as
\begin{align}
\mathbb{P}^{*}\left(\mathbb{P}\,\Psi\right)\,= & \,\,m_{0}^{2}c^{2}\Psi\,.\label{eq:15}
\end{align}
Using the quaternionic value of $\mathbb{P}^{*},\,\mathbb{P}$ and
$\Psi$, we get
\begin{align}
\mathbb{P}^{*}\left(\mathbb{P}\,\Psi\right)\,= & \,\,e_{0}\left\{ \frac{E^{2}}{c^{2}}-\left(p_{1}^{2}+p_{2}^{2}+p_{3}^{2}\right)\right\} \Psi_{0}+\sum_{j=1}^{3}e_{j}\left\{ \frac{E^{2}}{c^{2}}-\left(p_{1}^{2}+p_{2}^{2}+p_{3}^{2}\right)\right\} \Psi_{j}\,\,.\label{eq:16}
\end{align}
So, from Eq.(15) we obtain four relativistic quaternionic equations:
\begin{align}
E^{2}\Psi_{0}\,= & \,\,\left(p_{1}^{2}+p_{2}^{2}+p_{3}^{2}\right)c^{2}\Psi_{0}+m_{0}^{2}c^{4}\Psi_{0}\,,\,\,\,\,\,\,\,\,\,\,\,\,(\text{coefficient of }e_{0})\label{eq:17}\\
E^{2}\Psi_{j}\,= & \,\,\left(p_{1}^{2}+p_{2}^{2}+p_{3}^{2}\right)c^{2}\Psi_{j}+m_{0}^{2}c^{4}\Psi_{j}\,,\,\,\,\,\,\,\,\,\,\,\,\,(\text{coefficient of }e_{j})\,\,.\label{eq:18}
\end{align}
Now, to check these equations in quaternionic space-time frame, we
put operators $E=i\hbar\frac{\partial}{\partial t}$ and $p=-i\hbar\frac{\partial}{\partial x}$,
then the generalized quaternionic wave equations will be written as
\begin{align}
e_{0}\left(\boxdot-\frac{m_{0}^{2}c^{2}}{\text{\ensuremath{\hbar}}^{2}}\right)\Psi_{0}\,= & \,\,0\,,\,\,\,\,\,\,\,(\text{along scalar field)}\,;\label{eq:19}\\
e_{j}\left(\boxdot-\frac{m_{0}^{2}c^{2}}{\text{\ensuremath{\hbar}}^{2}}\right)\Psi_{j}\,= & \,\,0\,,\,\,\,\,\,\,\,(\text{along vector field)}\,;\label{eq:20}
\end{align}
where $\boxdot=\left(\nabla^{2}-\frac{1}{c^{2}}\frac{\partial^{2}}{\partial t^{2}}\right)$
represents the d' Alembert operator. Eqs.(19) and (20) represent the
quaternionic Klein-Gordon equation, respectively for scalar field
corresponding to scalar coefficient $e_{0}$ and vector field corresponding
to coefficient $e_{j}$. It is interesting to note that we may claim
that the generalized quaternionic field can unify relativistic Klein-Gordon
wave equations for the scalar and vector fields in a single frame,
such as
\begin{alignat}{1}
e_{0}\left(\boxdot-\kappa^{2}\right)\Psi_{0}+\sum_{j=1}^{3}e_{j}\left(\boxdot-\kappa^{2}\right)\Psi_{j} & \,\,=\,0\,,\,\,\,\,\,\,\,\,\,(\kappa=m_{0}c/\hbar)\,.\label{eq:21}
\end{alignat}
The generalized quaternionic Klein-Gordon equation is applicable if
we have a set of four electric or magnetic potentials that contain
both scalar and vector potentials. Moreover, we may also obtain the
group and phase velocities using the quaternionic wave function (13),
which are, respectively, $v_{g}=c^{2}k/\omega$ and $v_{p}=\pm c.$
As a result, a relation $v_{g}\times v_{p}=c^{2}$ may be used to
express the relationship between group and phase velocity for the
relativistic particles.

\section{Generalized quaternionic continuity equation and the probability
densities}

In this section, we describe how the quaternionic scalar and vector
fields flow in a four-dimensional space-time frame. Thus, to formulate
the generalized continuity equation for a quaternionic relativistic
wave propagation, we may use the quaternionic Klein-Gordon equation
given in Eq.(21) as
\begin{align}
\left(\boxdot-\frac{m^{2}c^{2}}{\text{\ensuremath{\text{\ensuremath{\hbar}}}}^{2}}\right)\Psi\,\equiv\,\,\left(\boxdot-\frac{m^{2}c^{2}}{\text{\ensuremath{\text{\ensuremath{\hbar}}}}^{2}}\right)\left(e_{0}\Psi_{0}+\sum_{j=1}^{3}e_{j}\Psi_{j}\right)\,= & \,\,0\,\,,\label{eq:22}
\end{align}
and its quaternionic conjugate
\begin{align}
\left(\text{\ensuremath{\boxdot}}-\frac{m^{2}c^{2}}{\text{\ensuremath{\text{\ensuremath{\hbar}}}}^{2}}\right)\Psi^{*}\,\equiv\,\,\left(\boxdot-\frac{m^{2}c^{2}}{\text{\ensuremath{\text{\ensuremath{\hbar}}}}^{2}}\right)\left(e_{0}\Psi_{0}^{*}-\sum_{j=1}^{3}e_{j}\Psi_{j}^{*}\right)\,= & \,\,0\,\,.\label{eq:23}
\end{align}
Now, multiplying the left side of Eq.(22) by quaternionic $\Psi^{*}$
and Eq.(23) by $\Psi$ and then subtracting both equations gives us
the following simplified version:
\begin{align}
 & \frac{i\text{\ensuremath{\hbar}}}{2mc^{2}}\frac{\partial}{\partial t}\left\{ \left[\Psi_{0}\frac{\partial}{\partial t}\Psi_{0}^{*}-\Psi_{0}^{*}\frac{\partial}{\partial t}\Psi_{0}\right]+\left[\left(\sum_{j=1}^{3}e_{j}\Psi_{j}\right)\frac{\partial}{\partial t}\left(\sum_{j=1}^{3}e_{j}\Psi_{j}\right)^{*}-\left(\sum_{j=1}^{3}e_{j}\Psi_{j}\right)^{*}\frac{\partial}{\partial t}\left(\sum_{j=1}^{3}e_{j}\Psi_{j}\right)\right]\right\} \nonumber \\
 & +\frac{i\text{\text{\ensuremath{\hbar}}}}{2mc^{2}}\frac{\partial}{\partial t}\left\{ \left[\left(\sum_{j=1}^{3}e_{j}\Psi_{j}\right)\frac{\partial}{\partial t}\Psi_{0}^{*}-\Psi_{0}^{*}\frac{\partial}{\partial t}\left(\sum_{j=1}^{3}e_{j}\Psi_{j}\right)\right]+\left[\Psi_{0}\frac{\partial}{\partial t}\left(\sum_{j=1}^{3}e_{j}\Psi_{j}\right)^{*}-\left(\sum_{j=1}^{3}e_{j}\Psi_{j}\right)^{*}\frac{\partial}{\partial t}\Psi_{0}\right]\right\} \nonumber \\
 & +\frac{i\text{\text{\ensuremath{\hbar}}}}{2m}\boldsymbol{\nabla}\left\{ \left[\Psi_{0}^{*}\boldsymbol{\nabla}\Psi_{0}-\Psi_{0}\boldsymbol{\nabla}\Psi_{0}^{*}\right]+\left[\left(\sum_{j=1}^{3}e_{j}\Psi_{j}\right)^{*}\boldsymbol{\nabla}\left(\sum_{j=1}^{3}e_{j}\Psi_{j}\right)-\left(\sum_{j=1}^{3}e_{j}\Psi_{j}\right)\boldsymbol{\nabla}\left(\sum_{j=1}^{3}e_{j}\Psi_{j}\right)^{*}\right]\right\} \nonumber \\
 & +\frac{i\text{\text{\ensuremath{\hbar}}}}{2m}\boldsymbol{\nabla}\left\{ \left[\Psi_{0}^{*}\boldsymbol{\nabla}\left(\sum_{j=1}^{3}e_{j}\Psi_{j}\right)-\left(\sum_{j=1}^{3}e_{j}\Psi_{j}\right)\boldsymbol{\nabla}\Psi_{0}^{*}\right]+\left[\left(\sum_{j=1}^{3}e_{j}\Psi_{j}\right)^{*}\boldsymbol{\nabla}\Psi_{0}-\Psi_{0}\boldsymbol{\nabla}\left(\sum_{j=1}^{3}e_{j}\Psi_{j}\right)^{*}\right]\right\} =0\,.\label{eq:24}
\end{align}
Furthermore, to compare with quaternionic continuity-like equation,
we estimate the quaternionic probability density (qPD) and quaternionic
probability current density (qPCD) from Eq.(24) as follows:
\begin{align}
\rho_{\text{qPD}}(e_{0},e_{j})\,=\, & \,\,e_{0}\left[\frac{i\text{\text{\ensuremath{\hbar}}}}{2mc^{2}}\left(\Psi_{0}\frac{\partial}{\partial t}\Psi_{0}^{*}-\Psi_{0}^{*}\frac{\partial}{\partial t}\Psi_{0}\right)+\frac{i\text{\text{\ensuremath{\hbar}}}}{2mc^{2}}\left(\sum_{j=1}^{3}\Psi_{j}\frac{\partial}{\partial t}\Psi_{j}^{*}-\sum_{j=1}^{3}\Psi_{j}^{*}\frac{\partial}{\partial t}\Psi_{j}\right)\right]\nonumber \\
 & \,+\sum_{j=0}^{3}e_{j}\left[\frac{i\text{\text{\ensuremath{\hbar}}}}{2mc^{2}}\left(\Psi_{j}\frac{\partial}{\partial t}\Psi_{0}^{*}-\Psi_{0}^{*}\frac{\partial}{\partial t}\Psi_{j}\right)+\frac{i\text{\text{\ensuremath{\hbar}}}}{2mc^{2}}\left(\Psi_{j}^{*}\frac{\partial}{\partial t}\Psi_{0}-\Psi_{0}\frac{\partial}{\partial t}\Psi_{j}^{*}\right)\right]\,\,,\label{eq:25}\\
J_{\text{qPCD}}(e_{0},e_{j})\,= & \,\,e_{0}\left[\frac{i\text{\text{\ensuremath{\hbar}}}}{2m}\left(\Psi_{0}^{*}\boldsymbol{\nabla}\Psi_{0}-\Psi_{0}\boldsymbol{\nabla}\Psi_{0}^{*}\right)+\frac{i\text{\text{\ensuremath{\hbar}}}}{2m}\left(\sum_{j=1}^{3}\Psi_{j}^{*}\boldsymbol{\nabla}\Psi_{j}-\sum_{j=1}^{3}\Psi_{j}\boldsymbol{\nabla}\Psi_{j}^{*}\right)\right]\nonumber \\
 & \,+\sum_{j=0}^{3}e_{j}\left[\frac{i\text{\text{\ensuremath{\hbar}}}}{2m}\left(\Psi_{0}^{*}\boldsymbol{\nabla}\Psi_{j}-\Psi_{j}\boldsymbol{\nabla}\Psi_{0}^{*}\right)+\frac{i\text{\text{\ensuremath{\hbar}}}}{2m}\left(\Psi_{0}\boldsymbol{\nabla}\Psi_{j}^{*}-\Psi_{j}^{*}\boldsymbol{\nabla}\Psi_{0}\right)\right]\,\,.\label{eq:26}
\end{align}
It is noticed that, $\rho_{\text{qPD}}$ and $J_{\text{qPCD}}$ are
there self both the quaternionic probability densities functions along
quaternionic basis $e_{0}$ and $e_{j}.$ To interpret qPD and qPCD,
let us take the three cases as
\begin{align*}
\text{Case 1:\,\,\,\,\,} & \rho_{\text{qPD}}=\frac{i\text{\text{\ensuremath{\hbar}}}}{2mc^{2}}\left(\Psi_{0}\frac{\partial}{\partial t}\Psi_{0}^{*}-\Psi_{0}^{*}\frac{\partial}{\partial t}\Psi_{0}\right),\,\,\,\,\,\,J_{\text{qPCD}}=\frac{i\text{\text{\ensuremath{\hbar}}}}{2m}\left(\Psi_{0}^{*}\boldsymbol{\nabla}\Psi_{0}-\Psi_{0}\boldsymbol{\nabla}\Psi_{0}^{*}\right),\,\,\,\,\,\,(\text{for scalar quaternion field)}\\
\text{Case 2:\,\,\,\,\,} & \rho_{\text{qPD}}=\frac{i\text{\text{\ensuremath{\hbar}}}}{2mc^{2}}\sum_{j=1}^{3}\left(\Psi_{j}\frac{\partial}{\partial t}\Psi_{j}^{*}-\Psi_{j}^{*}\frac{\partial}{\partial t}\Psi_{j}\right),\,\,\,\,\,\,J_{\text{qPCD}}=\frac{i\text{\text{\ensuremath{\hbar}}}}{2m}\sum_{j=1}^{3}\left(\Psi_{j}^{*}\boldsymbol{\nabla}\Psi_{j}-\Psi_{j}\boldsymbol{\nabla}\Psi_{j}^{*}\right),\,\,\,\,(\text{for pure quaternion field)}\\
\text{Case 3:\,\,\,\,\,} & \begin{cases}
\begin{array}{cc}
\rho_{\text{qPD}}= & \frac{i\text{\text{\ensuremath{\hbar}}}}{2mc^{2}}\sum_{j=1}^{3}\left[\left(\Psi_{j}\frac{\partial}{\partial t}\Psi_{0}^{*}-\Psi_{0}^{*}\frac{\partial}{\partial t}\Psi_{j}\right)+\left(\Psi_{j}^{*}\frac{\partial}{\partial t}\Psi_{0}-\Psi_{0}\frac{\partial}{\partial t}\Psi_{j}^{*}\right)\right]\,,\\
J_{\text{qPCD}}= & \frac{i\text{\text{\ensuremath{\hbar}}}}{2m}\sum_{j=1}^{3}\left[\left(\Psi_{0}^{*}\nabla\Psi_{j}-\Psi_{j}\nabla\Psi_{0}^{*}\right)+\left(\Psi_{0}\nabla\Psi_{j}^{*}-\Psi_{j}^{*}\nabla\Psi_{0}\right)\right]\,,
\end{array} & (\text{for combined quaternion field)}\end{cases}
\end{align*}
In the above instance, case 1 illustrates the conventional representation
of the probability density and probability current density of a relativistic
particle, but case 2 and case 3 take into account the probability
density and probability current density due to the effect of quaternionic
fields. The probability density and probability current density of
a pure quaternionic field depend solely on the vector fields, whereas
for the combined quaternionic field, they are influenced by the interactions
of scalar and vector fields. A somewhat similar outcome has previously
been suggested {[}12{]}.

\section{Quaternionic relativistic step-potential and the Klein Paradox}

In this section, we will discuss the application of the quaternionic
Klein-Gordon wave to the interaction with a quaternionic step potential.
We extend the idea of Giardino {[}8-11{]} where the quaternionic quantum
wave equation has been studied for a particle in a relativistic box.
Thus, let us consider a relativistic quaternionic step-potential $\mathbb{V}(\mathbb{X}):\rightarrow\mathbb{V}(t,\,x,y,z)$
or $\mathbb{V}(x,y,z,\,t)$ in four dimensional structure as
\begin{align}
\mathbb{V}(x,y,z,t)\,\,= & \,\,\begin{cases}
0\,\,, & \,\,\text{for\,\,\,}\mathbb{X}<0\,\\
\mathbb{V}_{0}\,\,, & \,\,\text{for\,}\,\,\mathbb{X}\geq0\,\,\,.
\end{cases}\label{eq:27}
\end{align}

\begin{figure}[H]
\noindent\begin{minipage}[t]{1\columnwidth}%
\begin{center}
\includegraphics[scale=0.5]{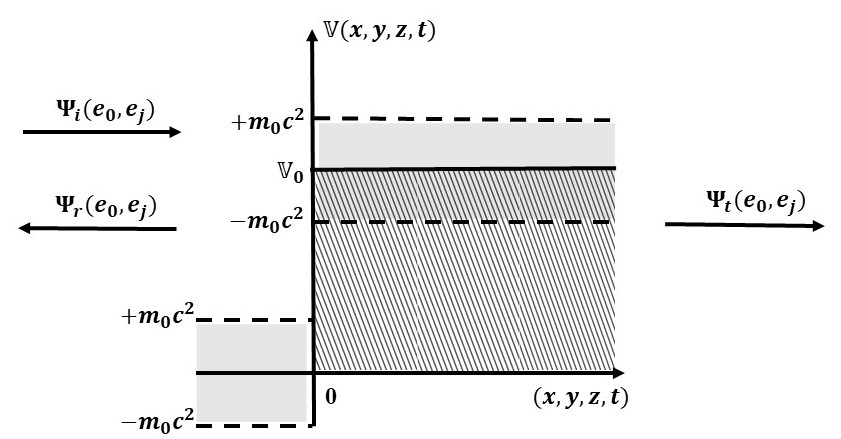}
\par\end{center}%
\end{minipage}\caption{Schematic diagram of quaternionic step potential}
\end{figure}
A relativistic quaternionic Klein-Gordon equation for the step potential
of regions $\mathbb{X}<0$ and $\mathbb{X}\geq0$ shown in Fig. 1
takes form,
\begin{align}
\left[-\frac{1}{c^{2}}\frac{\partial^{2}}{\partial t^{2}}+\nabla^{2}-\frac{m^{2}c^{2}}{\text{\ensuremath{\hbar}}^{2}}\right]\Psi(e_{0},e_{j})\,= & \,\,0\,,\,\,\,\,\,\,\,\,\,\,(\mathbb{X}<0)\,,\label{eq:28}\\
\left[\left(\frac{i}{c}\frac{\partial}{\partial t}-\mathbb{V}_{0}\right)^{2}+\nabla^{2}-\frac{m^{2}c^{2}}{\text{\text{\ensuremath{\hbar}}}^{2}}\right]\Psi(e_{0},e_{j})\,= & \,\,0\,,\,\,\,\,\,\,\,\,\,\,(\mathbb{X}\geq0)\,\,.\label{eq:29}
\end{align}
The quaternionic wave function is now defined as
\begin{align}
\Psi_{\text{\ensuremath{\mathbb{X}}\ensuremath{<0}}}\,\simeq & \,\,\Psi_{\mathcal{I}}+\Psi_{\mathcal{R}}\,,\,\,\,\,\,\,\,\,\,\,(\mathbb{X}<0)\label{eq:30}
\end{align}
where $\Psi_{\mathcal{I}}$ is the quaternionic form of incident wave
and $\Psi_{\mathcal{R}}$ is the quaternionic form of reflected wave,
which can be further expressed as,
\begin{align}
\Psi_{\mathcal{I}}(e_{0},e_{j})\,= & \,\,\exp\left[-\frac{i}{\text{\ensuremath{\hbar}}}\left(Et-\boldsymbol{p}\cdot\boldsymbol{x}\right)\right]+\sum_{j=1}^{3}e_{j}\exp\left[-\frac{i}{\text{\ensuremath{\hbar}}}\left(-ict\boldsymbol{p}+\frac{iE}{c}\boldsymbol{x}-\left(\boldsymbol{p}\times\boldsymbol{x}\right)_{j}\right)\right]\,\,,\label{eq:31}
\end{align}
and
\begin{align}
\Psi_{\mathcal{R}}(e_{0},e_{j})\,= & \,\,\mathcal{R}\exp\left[-\frac{i}{\text{\text{\ensuremath{\hbar}}}}\left(Et+\boldsymbol{p}\cdot\boldsymbol{x}\right)\right]+\sum_{j=1}^{3}e_{j}\mathcal{R}\exp\left[-\frac{i}{\text{\text{\ensuremath{\hbar}}}}\left(-ict\boldsymbol{p}-\frac{iE}{\text{\ensuremath{c}}}\boldsymbol{x}+\left(\boldsymbol{p}\times\boldsymbol{x}\right)_{j}\right)\right]\,\,,\label{eq:32}
\end{align}
where $\mathcal{R}$ represents the amplitude for the reflected wave.
As such, the quaternionic wave function for $\mathbb{X}\geq0$ becomes
$\Psi_{\text{\ensuremath{\mathbb{X}\geq0}}}=\Psi_{\mathcal{T}}$,
i.e.,
\begin{align}
\Psi_{\mathcal{T}}(e_{0},e_{j})\,= & \,\,\mathcal{T}\exp\left[-\frac{i}{\text{\text{\ensuremath{\hbar}}}}\left(Et-\boldsymbol{p}'\cdot\boldsymbol{x}\right)\right]+\sum_{j=1}^{3}e_{j}\mathcal{T}\exp\left[-\frac{i}{\text{\text{\ensuremath{\hbar}}}}\left(-ict\boldsymbol{p}'+\frac{iE}{\text{\text{\ensuremath{c}}}}\boldsymbol{x}-\left(\boldsymbol{p}'\times\boldsymbol{x}\right)_{j}\right)\right],\,\,\,(\mathbb{X}\geq0)\,,\label{eq:33}
\end{align}
with transmission amplitude $\mathcal{T}$ and momentum $\left|\boldsymbol{p'}\right|=\,\sqrt{\frac{1}{c^{2}}\left(E-\mathbb{V}_{0}\right)^{2}-m_{0}^{2}c^{2}}$.
Here, it is noticed that the kinetic energy of the transmitted particle
will be lesser than that of the incident kinetic energy ($E$) because
of the effect of step potential. Now, to solve the reflection and
transmission coefficients for the relativistic step potential, we
first calculate the reflected and transmitted amplitude of the quaternionic
wave function by applying boundary conditions. Let us consider the
boundary conditions
\begin{align}
\left.\Psi_{\text{\ensuremath{\mathbb{X}}\ensuremath{<0}}}\right|_{\left(0,\,t\right)}\,= & \,\,\left.\Psi_{\text{\ensuremath{\mathbb{X}\geq0}}}\right|_{\left(0,\,t\right)}\,\,;\label{eq:34}\\
\left.\dot{\Psi}_{\text{\ensuremath{\mathbb{X}}\ensuremath{<0}}}\right|_{\left(0,\,t\right)}\,= & \,\,\left.\dot{\Psi}_{\text{\ensuremath{\mathbb{X}\geq0}}}\right|_{\left(0,\,t\right)}\,\,,\label{eq:35}
\end{align}
which gives
\begin{align}
1+\mathcal{R} & \,=\,\,\frac{\exp\left[-\frac{i}{\text{\text{\ensuremath{\hbar}}}}Et\right]+\sum_{j=1}^{3}e_{j}\exp\left[-\frac{ct}{\text{\ensuremath{\hbar}}}\boldsymbol{p}'_{j}\right]}{\exp\left[-\frac{i}{\text{\text{\ensuremath{\hbar}}}}Et\right]+\sum_{j=1}^{3}e_{j}\exp\left[-\frac{ct}{\text{\ensuremath{\hbar}}}\boldsymbol{p}_{j}\right]}\mathcal{T}\,;\label{eq:36}\\
1-\mathcal{R} & \,=\,\,\frac{\frac{i}{\text{\text{\ensuremath{\hbar}}}}\boldsymbol{p}'\exp\left[-\frac{i}{\text{\text{\ensuremath{\hbar}}}}Et\right]+\sum_{j=1}^{3}\frac{E}{\text{\text{\ensuremath{\hbar}}}c}e_{j}\exp\left[-\frac{ct}{\text{\text{\ensuremath{\hbar}}}}\boldsymbol{p}'_{j}\right]}{\frac{i}{\text{\text{\ensuremath{\hbar}}}}\boldsymbol{p}\exp\left[-\frac{i}{\text{\text{\ensuremath{\hbar}}}}Et\right]+\sum_{j=1}^{3}\frac{E}{\text{\text{\ensuremath{\hbar}}}c}e_{j}\exp\left[-\frac{ct}{\text{\text{\ensuremath{\hbar}}}}\boldsymbol{p}_{j}\right]}\mathcal{T}\,,\label{eq:37}
\end{align}
dividing (36) by (37), then simplifying, we get
\begin{align}
|\mathcal{R}|^{2}\,= & \mathcal{\,\,R}\mathcal{R}^{*}\nonumber \\
= & \,\frac{\left(\boldsymbol{p}-\boldsymbol{p}'\right)^{2}+\left\{ \boldsymbol{p}e^{\left[-\frac{ct}{\text{\text{\ensuremath{\hbar}}}}\boldsymbol{p}'\right]}-\boldsymbol{p}'e^{\left[-\frac{ct}{\text{\text{\ensuremath{\hbar}}}}\boldsymbol{p}\right]}\right\} ^{2}+\frac{E^{2}}{c^{2}}\left\{ e^{\left[-\frac{ct}{\text{\text{\ensuremath{\hbar}}}}\boldsymbol{p}'\right]}-e^{\left[-\frac{ct}{\text{\text{\ensuremath{\hbar}}}}\boldsymbol{p}\right]}\right\} ^{2}}{\left(\boldsymbol{p}+\boldsymbol{p}'\right)^{2}+\frac{4E^{2}}{c^{2}}e^{\left[-\frac{2ct}{\text{\text{\ensuremath{\hbar}}}}\left(\boldsymbol{p}+\boldsymbol{p}'\right)\right]}+\left\{ \boldsymbol{p}e^{\left[-\frac{ct}{\text{\text{\ensuremath{\hbar}}}}\boldsymbol{p}'\right]}+\boldsymbol{p}'e^{\left[-\frac{ct}{\text{\text{\ensuremath{\hbar}}}}\boldsymbol{p}\right]}\right\} ^{2}+\frac{E^{2}}{c^{2}}\left\{ e^{\left[-\frac{ct}{\text{\text{\ensuremath{\hbar}}}}\boldsymbol{p}'\right]}+e^{\left[-\frac{ct}{\text{\text{\ensuremath{\hbar}}}}\boldsymbol{p}\right]}\right\} ^{2}}\,.\label{eq:38}
\end{align}
Here, $|\mathcal{R}|^{2}$ is the square of the amplitude of quaternionic
reflected wave. Now, the quaternionic reflection coefficient (qRC)
can be defined as
\begin{align}
\mathbb{R}(e_{0},e_{j})=- & \frac{J_{\mathcal{R}}}{J_{\mathcal{I}}}\,\,,\label{eq:39}
\end{align}
where $J_{\mathcal{R}}$and $J_{\mathcal{I}}$ are respectively the
reflected and the incident probability current densities that can
be determined by Eq.(26), such as
\begin{align}
J_{\mathcal{R}}\,= & \,\,\frac{\boldsymbol{p}}{m}|\mathcal{R}|^{2}+\frac{i\text{\ensuremath{\hbar}}}{2m}\sum_{j=1}^{3}e_{j}\left\llbracket \left(\frac{i\boldsymbol{p}}{\text{\text{\ensuremath{\hbar}}}}-\frac{E}{\text{\text{\ensuremath{\hbar}}}c}\right)\exp\left\{ -\frac{i}{\text{\text{\ensuremath{\hbar}}}}\left[\left(Et+\boldsymbol{p}\cdot\boldsymbol{x}\right)-\left(ict\boldsymbol{p}+\frac{iE}{\text{\text{\ensuremath{c}}}}\boldsymbol{x}+\left(\boldsymbol{p}\times\boldsymbol{x}\right)_{j}\right)\right]\right\} \right.\nonumber \\
 & \left.-\left(\frac{i\boldsymbol{p}}{\text{\text{\ensuremath{\hbar}}}}+\frac{E}{\text{\text{\ensuremath{\hbar}}}c}\right)\exp\left\{ \frac{i}{\text{\text{\ensuremath{\hbar}}}}\left[\left(Et+\boldsymbol{p}\cdot\boldsymbol{x}\right)-\left(-ict\boldsymbol{p}-\frac{iE}{\text{\text{\ensuremath{c}}}}\boldsymbol{x}+\left(\boldsymbol{p}\times\boldsymbol{x}\right)_{j}\right)\right]\right\} \right\rrbracket |\mathcal{R}|^{2}\,\,,\label{eq:40}
\end{align}
 and
\begin{align}
J_{\mathcal{I}}\,= & \,-\frac{\boldsymbol{p}}{m}+\frac{i\text{\text{\ensuremath{\hbar}}}}{2m}\sum_{j=1}^{3}e_{j}\left\llbracket \left(\frac{i\boldsymbol{p}}{\text{\text{\ensuremath{\hbar}}}}+\frac{E}{\text{\text{\ensuremath{\hbar}}}c}\right)\exp\left\{ \frac{i}{\text{\text{\ensuremath{\hbar}}}}\left[\left(Et-\boldsymbol{p}\cdot\boldsymbol{x}\right)-\left(-ict\boldsymbol{p}+\frac{iE}{\text{\ensuremath{c}}}\boldsymbol{x}-\left(\boldsymbol{p}\times\boldsymbol{x}\right)_{j}\right)\right]\right\} \right.\nonumber \\
 & \left.-\left(\frac{i\boldsymbol{p}}{\text{\text{\ensuremath{\hbar}}}}-\frac{E}{\text{\text{\ensuremath{\hbar}}}c}\right)\exp\left\{ -\frac{i}{\text{\text{\ensuremath{\hbar}}}}\left[\left(Et-\boldsymbol{p}\cdot\boldsymbol{x}\right)-\left(ict\boldsymbol{p}-\frac{iE}{\text{\ensuremath{c}}}\boldsymbol{x}-\left(\boldsymbol{p}\times\boldsymbol{x}\right)_{j}\right)\right]\right\} \right\rrbracket \,\,,\label{eq:41}
\end{align}
Thus, from Eq.(39) the quaternionic reflection coefficient becomes
\begin{align}
 & \mathbb{R}(e_{0},e_{j})\,=\nonumber \\
 & -\frac{\frac{\boldsymbol{p}}{m}+\frac{i\text{\text{\ensuremath{\hbar}}}}{2m}\sum_{j=1}^{3}e_{j}\left[\left(\frac{i\boldsymbol{p}}{\text{\text{\ensuremath{\hbar}}}}-\frac{E}{\text{\text{\ensuremath{\hbar}}}c}\right)e^{-\frac{i}{\text{\text{\ensuremath{\hbar}}}}\left[Et+\boldsymbol{p}\cdot\boldsymbol{x}-ict\boldsymbol{p}-\frac{iE}{\text{\text{\ensuremath{c}}}}\boldsymbol{x}-\left(\boldsymbol{p}\times\boldsymbol{x}\right)_{j}\right]}-\left(\frac{i\boldsymbol{p}}{\text{\text{\ensuremath{\hbar}}}}+\frac{E}{\text{\text{\ensuremath{\hbar}}}c}\right)e^{\frac{i}{\text{\text{\ensuremath{\hbar}}}}\left[Et+\boldsymbol{p}\cdot\boldsymbol{x}+ict\boldsymbol{p}+\frac{iE}{\text{\text{\ensuremath{c}}}}\boldsymbol{x}-\left(\boldsymbol{p}\times\boldsymbol{x}\right)_{j}\right]}\right]}{-\frac{\boldsymbol{p}}{m}+\frac{i\text{\text{\ensuremath{\hbar}}}}{2m}\sum_{j=1}^{3}e_{j}\left[\left(\frac{i\boldsymbol{p}}{\text{\text{\ensuremath{\hbar}}}}+\frac{E}{\text{\text{\ensuremath{\hbar}}}c}\right)e^{\frac{i}{\text{\text{\ensuremath{\hbar}}}}\left[Et-\boldsymbol{p}\cdot\boldsymbol{x}+ict\boldsymbol{p}-\frac{iE}{\text{\ensuremath{c}}}\boldsymbol{x}+\left(\boldsymbol{p}\times\boldsymbol{x}\right)_{j}\right]}-\left(\frac{i\boldsymbol{p}}{\text{\text{\ensuremath{\hbar}}}}-\frac{E}{\text{\text{\ensuremath{\hbar}}}c}\right)e^{-\frac{i}{\text{\text{\ensuremath{\hbar}}}}\left[Et-\boldsymbol{p}\cdot\boldsymbol{x}-ict\boldsymbol{p}+\frac{iE}{\text{\ensuremath{c}}}\boldsymbol{x}+\left(\boldsymbol{p}\times\boldsymbol{x}\right)_{j}\right]}\right]}|\mathcal{R}|^{2}\,,\label{eq:42}
\end{align}
We can split the quaternionic reflection coefficient into two parts,
where the pure scalar part yields
\begin{align}
\mathbb{R}(e_{0},0) & \,\,=\,|\mathcal{R}|^{2}\,,\label{eq:43}
\end{align}
while the pure quaternionic part becomes
\begin{align}
\mathbb{R}(0,e_{j})\, & =\,\,\sum_{j=1}^{3}e_{j}\left(\frac{\frac{E}{c}\cos\alpha-\boldsymbol{p}\sin\alpha}{\frac{E}{c}\cos\beta-\boldsymbol{p}\sin\beta}\right)\exp\left[-\frac{2E}{\text{\ensuremath{\hbar}}c}\boldsymbol{x}_{j}\right]|\mathcal{R}|^{2}\,,\label{eq:44}
\end{align}
where the parameters $\alpha=\frac{1}{\text{\ensuremath{\hbar}}}\left[Et+\boldsymbol{p}\cdot\boldsymbol{x}-\left(\boldsymbol{p}\times\boldsymbol{x}\right)\right]$
and $\beta=\frac{1}{\text{\ensuremath{\hbar}}}\left[Et-\boldsymbol{p}\cdot\boldsymbol{x}+\left(\boldsymbol{p}\times\boldsymbol{x}\right)\right]$.
Therefore, it is conclude that the real quaternionic reflection coefficient
has conventional meaning, however, the pure quaternionic reflection
coefficient shows the additional part associated with quaternionic
phase part and the decaying factor $\Omega(\boldsymbol{x}_{j})\,\sim\,\exp\left[-\frac{2E}{\text{\ensuremath{\hbar}}c}\boldsymbol{x}_{j}\right]$
along with $e_{j}-$ frames arise from the quaternionic wave function.
\\
Besides, the quaternionic transmission coefficient (qTC) can be identified
as
\begin{align}
\mathbb{T}(e_{0},e_{j})\,= & \,\,\frac{J_{\mathcal{T}}}{J_{\mathcal{I}}}\,\,,\label{eq:45}
\end{align}
where
\begin{align}
J_{\mathcal{T}}\,= & \,-\frac{\boldsymbol{p}'}{m}|\mathcal{T}|^{2}+\frac{i\text{\text{\ensuremath{\hbar}}}}{2m}\sum_{j=1}^{3}e_{j}\left\llbracket \left(\frac{i\boldsymbol{p}'}{\text{\text{\ensuremath{\hbar}}}}+\frac{E}{\text{\text{\ensuremath{\hbar}}}c}\right)\exp\frac{i}{\text{\text{\ensuremath{\hbar}}}}\left[\left(Et-\boldsymbol{p}'\cdot\boldsymbol{x}\right)-\left(-ict\boldsymbol{p}'+\frac{iE}{\text{\ensuremath{c}}}\boldsymbol{x}-\left(\boldsymbol{p}'\times\boldsymbol{x}\right)_{j}\right)\right]\right.\nonumber \\
 & \left.-\left(\frac{i}{\text{\text{\ensuremath{\hbar}}}}\boldsymbol{p}'-\frac{E}{\text{\text{\ensuremath{\hbar}}}c}\right)\exp\left\{ -\frac{i}{\text{\text{\ensuremath{\hbar}}}}\left[\left(Et-\boldsymbol{p}'\cdot\boldsymbol{x}\right)-\left(ict\boldsymbol{p}'-\frac{iE}{\text{\ensuremath{c}}}\boldsymbol{x}-\left(\boldsymbol{p}'\times\boldsymbol{x}\right)_{j}\right)\right]\right\} \right\rrbracket \,|\mathcal{T}|^{2}\,\,.\label{eq:46}
\end{align}
Therefore, by substituting the value of $J_{\mathcal{T}}$ and $J_{\mathcal{I}},$
the quaternionic transmission coefficient becomes
\begin{align}
 & \mathbb{T}(e_{0},e_{j})\,=\nonumber \\
 & \frac{-\frac{\boldsymbol{p}'}{m}+\frac{i\text{\text{\ensuremath{\hbar}}}}{2m}\sum_{j=1}^{3}e_{j}\left[\left(\frac{i\boldsymbol{p}'}{\text{\text{\ensuremath{\hbar}}}}+\frac{E}{\text{\text{\ensuremath{\hbar}}}c}\right)e^{\frac{i}{\text{\text{\ensuremath{\hbar}}}}\left[Et-\boldsymbol{p}'\cdot\boldsymbol{x}+ict\boldsymbol{p}'-\frac{iE}{\text{\ensuremath{c}}}\boldsymbol{x}+\left(\boldsymbol{p}'\times\boldsymbol{x}\right)_{j}\right]}-\left(\frac{i\boldsymbol{p}'}{\text{\text{\ensuremath{\hbar}}}}-\frac{E}{\text{\text{\ensuremath{\hbar}}}c}\right)e^{-\frac{i}{\text{\text{\ensuremath{\hbar}}}}\left[Et-\boldsymbol{p}'\cdot\boldsymbol{x}-ict\boldsymbol{p}'+\frac{iE}{\text{\ensuremath{c}}}\boldsymbol{x}+\left(\boldsymbol{p}'\times\boldsymbol{x}\right)_{j}\right]}\right]}{-\frac{\boldsymbol{p}}{m}+\frac{i\text{\text{\ensuremath{\hbar}}}}{2m}\sum_{j=1}^{3}e_{j}\left[\left(\frac{i\boldsymbol{p}}{\text{\text{\ensuremath{\hbar}}}}+\frac{E}{\text{\text{\ensuremath{\hbar}}}c}\right)e^{\frac{i}{\text{\text{\ensuremath{\hbar}}}}\left[Et-\boldsymbol{p}\cdot\boldsymbol{x}+ict\boldsymbol{p}-\frac{iE}{\text{\ensuremath{c}}}\boldsymbol{x}+\left(\boldsymbol{p}\times\boldsymbol{x}\right)_{j}\right]}-\left(\frac{i\boldsymbol{p}}{\text{\text{\ensuremath{\hbar}}}}-\frac{E}{\text{\text{\ensuremath{\hbar}}}c}\right)e^{-\frac{i}{\text{\text{\ensuremath{\hbar}}}}\left[Et-\boldsymbol{p}\cdot\boldsymbol{x}-ict\boldsymbol{p}+\frac{iE}{\text{\ensuremath{c}}}\boldsymbol{x}+\left(\boldsymbol{p}\times\boldsymbol{x}\right)_{j}\right]}\right]}\,|\mathcal{T}|^{2}.\label{eq:47}
\end{align}
The real part of quaternionic transmission coefficient will be contributed
as
\begin{align}
\mathbb{T}(e_{0},0)\,= & \,\,|\mathcal{T}|^{2}\,\,,\label{eq:48}
\end{align}
while the pure quaternionic part contributes
\begin{align}
\mathbb{T}(0,e_{j})\,= & \,\,\sum_{j=1}^{3}e_{j}\left(\frac{\frac{E}{c}\cos\alpha'-\boldsymbol{p}'\sin\alpha'}{\frac{E}{c}\cos\beta'-\boldsymbol{p}\sin\beta'}\right)\exp\left[-ct\left(\boldsymbol{p}'-\boldsymbol{p}\right)\right]\,|\mathcal{T}|^{2}\,\,,\label{eq:49}
\end{align}
where $\alpha'\,=\,\frac{1}{\hbar}\left[Et-\boldsymbol{p'}\cdot\boldsymbol{x}+\left(\boldsymbol{p}'\times\boldsymbol{x}\right)_{j}\right]$
and $\beta'\,=\,\frac{1}{\hbar}\left[Et-\boldsymbol{p}\cdot\boldsymbol{x}+\left(\boldsymbol{p}\times\boldsymbol{x}\right)_{j}\right]$.
Here, the real quaternionic transmission coefficient has usual form,
but the pure quaternionic transmission coefficient has an additional
part associated with the decaying factor $\exp\left[-ct\left(\boldsymbol{p}'-\boldsymbol{p}\right)\right]$
for $e_{j}-$ axes arise from the quaternionic wave function. Further,
if particles are not to be created or destroyed at the boundary as
in a real quaternions, it strongly follows $\mathbb{R}+\mathbb{T}=1$,
but for pure quaternionic case, the relation violates due to relativistic
destroy in quaternionic space. Now, to check the significance of the
quaternionic formalism, in following subsections we will discuss some
important cases for the incident kinetic energy of particle that compare
to the quaternionic step potential.

\subsection{$\boldsymbol{E>\mathbb{V}_{0}+m_{0}c^{2}}$ : The oscillatory-like
zone}

In this case the kinetic energy of particle is greater than the sum
of the step-potential and rest mass energy, then the momentum of the
particle for region $\mathbb{X}\geq0$ will be positive, i.e., $\left.\boldsymbol{P}'\right|_{\mathbb{X}\geq0}=\boldsymbol{p}'$.
Here, we assume that the lowest feasible value of step potential becomes
the rest mass energy of particle. In order to check the quaternionic
significance with time varying expression, we restrict our solutions
only for time slaps $t\rightarrow0$ and $t\rightarrow\infty$. 
\begin{itemize}
\item \textbf{Case 1: $t\rightarrow0$}. For this case, the square of reflection
and transmission amplitude become:
\begin{align}
|\mathcal{R}|^{2}\,= & \,\,\frac{\left(p-p'\right)^{2}}{\left(p+p'\right)^{2}-4p_{0}^{2}}\,\,\,;\,\,\,\,\,\,|\mathcal{T}|^{2}\,=\,\,\frac{4pp'-4p_{0}^{2}}{\left(p+p'\right)^{2}-4p_{0}^{2}}\,.\label{eq:50}
\end{align}
Thus, the pure quaternionic form of reflection and transmission coefficients
can be expressed by
\begin{align}
\mathbb{R}(e_{0},0)\,= & \,\,|\mathcal{R}|^{2}\,,\,\,\,\,\mathbb{R}(0,e_{j})\,=\,\left(\frac{\frac{E}{c}\cos\alpha_{0}-\boldsymbol{p}\sin\alpha_{0}}{\frac{E}{c}\cos\beta_{0}-\boldsymbol{p}\sin\beta_{0}}\right)\Omega(\boldsymbol{x}_{j})|\mathcal{R}|^{2}\,;\label{eq:51}\\
\mathbb{T}(e_{0},0)\,= & \,\,|\mathcal{T}|^{2}\,,\,\,\,\,\mathbb{T}(0,e_{j})\,=\,\left(\frac{\frac{E}{c}\cos\alpha_{0}'-\boldsymbol{p}'\sin\alpha_{0}'}{\frac{E}{c}\cos\beta_{0}-\boldsymbol{p}\sin\beta_{0}}\right)|\mathcal{T}|^{2}\,,\label{eq:52}
\end{align}
where $\alpha_{0}=\frac{1}{\text{\ensuremath{\hbar}}}\left[\boldsymbol{p}\cdot\boldsymbol{x}-\left(\boldsymbol{p}\times\boldsymbol{x}\right)_{j}\right]$,
$\beta_{0}=\frac{1}{\hbar}\left[-\boldsymbol{p}\cdot\boldsymbol{x}+\left(\boldsymbol{p}\times\boldsymbol{x}\right)_{j}\right]$and
$\alpha_{0}'=\frac{1}{\hbar}\left[-\boldsymbol{p'}\cdot\boldsymbol{x}+\left(\boldsymbol{p}'\times\boldsymbol{x}\right)_{j}\right]$.
The reflection and transmission probability of a particle at the boundary
of a quaternionic step potential is prevalent in pure scalar axes;
however, it has some significant consequences in pure quaternionic
axes. Despite the fact that the kinetic energy of a relativistic particle
is greater than the quaternionic step potential, the probability of
some particles reflecting back has a declining exponential component,
whereas transmission has an oscillatory form for pure quaternionic
space.
\item \textbf{Case 2:} $t\rightarrow\infty$. In this case, the step potential
width of forth-space component is taken extremely large. Then, the
quaternionic reflection and transmission coefficients become:
\begin{align}
\mathbb{R}(e_{0},0)\,= & \,\,\frac{\left(p-p'\right)^{2}}{\left(p+p'\right)^{2}}\,\,;\,\,\,\,\,\mathbb{R}(0,e_{j})\,=\,\Omega(\boldsymbol{x}_{j})|\mathcal{R}|^{2}\,\,,\label{eq:53}\\
\mathbb{T}(e_{0},0)\,= & \,\,\mathbb{T}(0,e_{j})\,=\,0\,.\label{eq:54}
\end{align}
It is interesting to note that in this case, the quaternionic usual
reflection coefficient is all that exists, and there is no quaternionic
transmission for the potential steps that result from the fourth space
component expanding towards infinity. The particle may therefore act
like a classical particle.
\end{itemize}

\subsection{$\boldsymbol{\left(\mathbb{V}_{0}-m_{0}c^{2}\right)<E<\left(\mathbb{V}_{0}+m_{0}c^{2}\right)}$:
The tunnelling-like zone}

In this case, the kinetic energy of the particle varies within the
quaternionic effective potential, i.e., from $\mathbb{V}_{0}-m_{0}c^{2}$
to $\mathbb{V}_{0}+m_{0}c^{2}$ (see Fig. 1). Then, for this energy
band, one can have
\begin{equation}
\left.\boldsymbol{P'}\right|_{\mathbb{X}\geq0}\,=\,i\boldsymbol{p'}\label{eq:55}
\end{equation}
where $\boldsymbol{P}'$ indicates the imaginary momentum of the particle
for region $\mathbb{X}\geq0$, so we can simply replace $i\boldsymbol{p'}$
instead of $\boldsymbol{p}$$'$.
\begin{itemize}
\item \textbf{Case 1: $t\rightarrow0$}. For this case, the square of reflection
amplitude and transmission amplitude modified as:
\begin{align}
|\mathcal{R}|^{2}\,= & \,\,\frac{\left(p-ip'\right)^{2}}{\left(p+ip'\right)^{2}-4p_{0}^{2}}\,\,;\,\,\,\,\,\,\,\,|\mathcal{T}|^{2}\,=\,\frac{4ipp'-4p_{0}^{2}}{\left(p+ip'\right)^{2}-4p_{0}^{2}}\,\,,\label{eq:56}
\end{align}
which gives
\begin{align}
\mathbb{R}(e_{0},0)\,= & \,\,|\mathcal{R}|^{2}\,;\,\,\,\,\mathbb{R}(0,e_{j})\,=\,\,\left(\frac{\frac{E}{c}\cos\alpha_{0}-\boldsymbol{p}\sin\alpha_{0}}{\frac{E}{c}\cos\beta_{0}-\boldsymbol{p}\sin\beta_{0}}\right)\Omega(\boldsymbol{x}_{j})|\mathcal{R}|^{2}\,,\label{eq:57}\\
\mathbb{T}(e_{0},0)\,= & \,\,|\mathcal{T}|^{2}\,;\,\,\,\,\mathbb{T}(0,e_{j})\,=\,\,\left(\frac{\frac{E}{c}\cos\alpha_{0}'-i\boldsymbol{p}'\sin\alpha_{0}'}{\frac{E}{c}\cos\beta_{0}-\boldsymbol{p}\sin\beta_{0}}\right)|\mathcal{T}|^{2}\,.\label{eq:58}
\end{align}
However, for time $t\rightarrow\tau$ the pure quaternionic transmission
coefficient varies as
\begin{align}
\left.\mathbb{T}(0,e_{j})\right|_{t\rightarrow\tau}\,=\,\,\mathbb{T}(0,e_{j})\, & e^{-\xi\,\tau}\,,\label{eq:59}
\end{align}
where $\xi=c\boldsymbol{p}'-c\boldsymbol{p}$ indicates the change
of relativistic kinetic energy of a moving particle from region $\mathbb{X}<0$
to $\mathbb{X}\geq0$. As a result of Eq.(59), the transmission coefficient
for a pure quaternionic framework decays exponentially with time.
Thus, the probability of identifying particles in $\mathbb{X}>0$
within such a tunnelling-like zone can be estimated by the fourth-space
component of pure quaternionic space.
\item \textbf{Case 2:} $t\rightarrow\infty$. The reflection and transmission
coefficients for the infinitely extended step potential width due
to the quaternionic fourth-space component are
\begin{align}
\mathbb{R}(e_{0},0)\,= & \,\,|\mathcal{R}|^{2}\,\,\equiv\,\,\frac{\left(p-ip'\right)^{2}}{\left(p+ip'\right)^{2}}\,\,\,;\,\,\,\,\,\mathbb{R}(0,e_{j})\,=\,\Omega(\boldsymbol{x}_{j})|\mathcal{R}|^{2}\,\,,\label{eq:60}\\
\mathbb{T}(e_{0},0)\,= & \,\,\mathbb{T}(0,e_{j})\,=\,0\,.\label{eq:61}
\end{align}
Essentially, as shown in subsection 6.2, only the quaternionic usual
reflection coefficient is present in this situation, and no quaternionic
transmission takes place for any hypothetical steps that result in
the fourth space component stretching towards infinity.
\end{itemize}

\subsection{$\boldsymbol{E<\left(\mathbb{V}_{0}-m_{0}c^{2}\right)}$: The Klein-like
zone}

This zone is crucial because if the particle's kinetic energy is significantly
lower than the height of the tunnelling-like zone, inappropriate particle
reflection will have appeared. In order to analyze the Klein paradox
behavior of relativistic particles in a quaternionic frame, let us
write the momentum of the particle for region $\mathbb{X}\geq0$,
which becomes negative as
\begin{equation}
\left.\boldsymbol{P}'\right|_{\mathbb{X}\geq0}=\,\,-\boldsymbol{p}'\,\,.\label{eq:62}
\end{equation}
To examine the probability of particles in the regions $\mathbb{X}<0$
and $\mathbb{X}\geq0$, we can place this opposite momentum value
in Eqs.(39) and (45) as the following manner:
\begin{itemize}
\item \textbf{Case 1:} $t\rightarrow0$. In this instance, the square of
the reflection and transmission amplitudes of the quaternionic wave
is adjusted as follows:
\begin{align}
|\mathcal{R}|^{2}\,\,=\,\,\frac{\left(p+p'\right)^{2}}{\left(p-p'\right)^{2}-4p_{0}^{2}}\,\,;\,\,\,\,\,\,\,\,\,|\mathcal{T}|^{2}\,\,= & \,\,\frac{-4\left(pp'+p_{0}^{2}\right)}{\left(p-p'\right)^{2}-4p_{0}^{2}}\,\,.\label{eq:63}
\end{align}
Hence, the quaternionic reflection coefficients i.e., $\mathbb{R}(e_{0},0),\,$$\mathbb{R}(0,e_{j})$
become:
\begin{align}
\mathbb{R}(e_{0},0)\,= & \,\,|\mathcal{R}|^{2}\,\,\,;\,\,\,\,\,\,\mathbb{R}(0,e_{j})\,=\,\,\left(\frac{\frac{E}{c}\cos\alpha_{0}-\boldsymbol{p}\sin\alpha_{0}}{\frac{E}{c}\cos\beta_{0}-\boldsymbol{p}\sin\beta_{0}}\right)\Omega(\boldsymbol{x}_{j})|\mathcal{R}|^{2}\,\,.\label{eq:64}
\end{align}
Since we see that $|\mathcal{R}|^{2}>1$, it is clear that the reflection
coefficient increases and exceeds value one in both pure scalar and
pure quaternionic circumstances that is providing an incompatible
outcome for the quaternionic Klein paradox. A reflection coefficient
larger than one caused by opposite momentum due to the quaternionic
quantum fluctuations implies more particles will reflect that fall
on it. However, the transmission coefficient changes to:
\begin{align}
\mathbb{T}(e_{0},0)\,= & \,\,|\mathcal{T}|^{2}\,\,\,;\,\,\,\,\,\,\mathbb{T}(0,e_{j})\,=\,\left[\frac{\frac{E}{c}\cos\alpha_{0}'+\boldsymbol{p}'\sin\alpha_{0}'}{\frac{E}{c}\cos\beta_{0}-\boldsymbol{p}\sin\beta_{0}}\right]|\mathcal{T}|^{2}\,\,.\label{eq:65}
\end{align}
The relevance of the quaternionic Klein paradox is provided by the
fact that $|\mathcal{T}|^{2}<0$, which means that the quaternionic
transmission coefficients are negative for both pure scalar and pure
quaternionic conditions.
\item \textbf{Case 2:} $t\rightarrow\infty$. For this situation, the reflection
and the transmission coefficients are expressed by
\begin{align}
\mathbb{R}(e_{0},0)\,= & \,\,|\mathcal{R}|^{2}\,\equiv\,\frac{\left(p+p'\right)^{2}}{\left(p-p'\right)^{2}}\,\,\,;\,\,\,\,\,\,\,\,\mathbb{R}(0,e_{j})\,=\,\Omega(\boldsymbol{x}_{j})|\mathcal{R}|^{2}\,\,,\label{eq:66}\\
\mathbb{T}(e_{0},0)\,= & \,\,\mathbb{T}(0,e_{j})=\,0\,.\label{eq:67}
\end{align}
Therefore, the probability of finding the particles in the reflected
case is still greater than one, as explained in case $t\rightarrow0$,
but it impacts zero for the quaternionic transmitted probability when
we increase the fourth dimension of space in the potential step up
to infinity. 
\end{itemize}
Hence, it is claimed that by reducing the step potential width at
a small scale, the fundamental Klein-Gordon wave equation in quaternionic
($3+1$)$-$fields may be extended to explain the dynamics of relativistic
particles in finite potential barrier.

\section{Conclusion}

The proposed research examines the Klein-Gordon wave equation in the
context of quaternionic step potential by employing the $(3+1)-$dimensional
quaternionic wave function. The quaternionic generalization of the
Klein-Gordon equation has been established for both scalar and vector
fields. We reviewed the quaternionic continuity-like equation that
connected probability density and probability current density for
pure scalars, pure vectors, and their mixed fields. It is intriguing
that the combined field results from the mutual interaction of the
real and pure quaternionic fields. Furthermore, the amplitudes of
the incident, reflected, and transmitted waves for the quaternionic
four-potential step have been analyzed. It is emphasized that the
pure scalar component of the quaternionic incident, reflected, and
transmitted waves is analogous to conventional waves; however, the
additional or pure vector component of these waves appears due to
the presence of a quaternionic structure. Correspondingly, we found
the reflection and transmission coefficients associated with the quaternionic
four-potential step, which demonstrate the probability of the existence
of waves (or particles) in different regions of the step potential
problem. Surprisingly, the quaternionic value of the reflected and
transmitted coefficients is determined by the particle's four-momentum.
Furthermore, we have addressed the quaternionic interpretation for
the oscillatory, tunnelling, and Klein paradox instances in which
the incoming wave-function interacts with the potential step via the quaternionic
momentum of the particle. It has also been claimed that the physical
conservation law $\mathbb{R}+\mathbb{T}=1$ does not hold for pure
quaternionic fields owing to non-commutativity, although it holds
for pure scalar fields. Subsequently, it has been decided that, in
order to account for the possibility of the quaternionic Klein paradox,
the particle momentum in the step potential region ($\mathbb{X}\geq0$)
must be opposite to that region ($\mathbb{X}<0$), which suggests
that the kinetic energy of the incoming wave (or particle) becomes
negative. For the quaternionic Klein paradox, the quaternionic reflection
coefficient becomes solely higher than one and the quaternionic transmission
coefficient becomes simply less than zero. As a result, the current
quaternionic ($3+1$) formalism is sufficiently consistent to explain
the relativistic quantum phenomena of the contemporary era.

\end{document}